\documentclass[prl, twocolumn]{revtex4}

\usepackage{amsmath}
\usepackage[english]{babel}

\def\tr{{\rm Tr}}

\begin{document}

\title{Adiabatic Markovian Dynamics}

\author{Ognyan Oreshkov$^{1,2}$ and John Calsamiglia$^1$}

\affiliation{${^1}$F\'isica Te\`orica: Informaci\'o i Fen\`omens Qu\`antics, Universitat
Aut\`{o}noma de Barcelona, 08193 Bellaterra (Barcelona), Spain\\
${^2}$Faculty of Physics, University of Vienna, Boltzmanngasse 5,
A-1090 Vienna, Austria}
\date{\today}

\begin{abstract}

We propose a theory of adiabaticity in quantum Markovian dynamics
based on a decomposition of the Hilbert space induced by
the asymptotic behavior of the Lindblad semigroup. A central idea of
our approach is that the natural generalization of the concept of
eigenspace of the Hamiltonian in the case of Markovian dynamics is a
noiseless subsystem with a minimal noisy cofactor. Unlike previous
attempts to define adiabaticity for open systems, our approach deals
exclusively with physical entities and provides a simple, intuitive
picture at the underlying Hilbert-space level, linking the notion of
adiabaticity to the theory of noiseless subsystems. As an
application of our theory, we propose a framework for
decoherence-assisted computation in noiseless codes under general
Markovian noise. We also formulate a dissipation-driven approach to
holonomic computation based on adiabatic dragging of subsystems that
is generally not achievable by non-dissipative means.

\end{abstract}

\maketitle



\textit{Introduction.}---The adiabatic theorem is a simple and
powerful result that has been known since the early days of quantum
mechanics \cite{BornFock,Kato50}. It states roughly that a closed
system in an eigenstate of a continuously perturbed Hamiltonian
remains in an instantaneous eigenstate in the limit of slow
perturbations if the corresponding eigenvalue is separated from the
rest of the spectrum by a gap. Adiabaticity in quantum mechanics has
applications in a wide range of areas, including quantum chemistry
\cite{BornOppenheimer}, geometric phases \cite{Berry84, WZ84},
quantum Hall effect \cite{Loughlin81}, STIRAP \cite{Vitanov}, and
quantum phase transitions \cite{Sachdev}. More recently, the
adiabatic theorem has been the subject of increased interest in
relation to quantum information processing, where it has served as a
basis for a variety of schemes, including holonomic quantum
computation \cite{ZR99} and adiabatic quantum algorithms
\cite{FGGS00}.

Given the importance of the concept of adiabaticity in closed
quantum systems, it is natural to ask how this concept extends to
the dynamics of systems interacting with an environment. This
question is of particular interest from the point of view of quantum
information processing where decoherence is a major obstacle to the
construction of reliable quantum devices, and at the same time
non-unitary processes are an important tool for quantum control
\cite{NieChu00}. In Ref.~\cite{SL05}, Sarandy and Lidar proposed an
approach to the adiabatic dynamics of open quantum systems, defining
adiabaticity as the regime in which the operator subspaces
corresponding to the instantaneous Jordan blocks of the generator of
the dynamics evolve independently (for adiabaticity in weakly open
systems, see Ref.~\cite{Thunstrom05}). This definition is motivated
by the formal analogy between the Schr\"{o}dinger equation and the
time-dependent Markovian master equation written in a coherence
basis, both being first-order linear vector differential equations
with the difference that the generator of the master equation is
generally not diagonalizable (hence the Jordan decomposition). But
while in closed systems the phenomenon of adiabaticity concerns the
decoupled evolution of eigenspaces of the Hamiltonian which
themselves are Hilbert spaces containing physical states, the Jordan
blocks correspond to generally nonorthogonal subspaces of the space
of linear operators that need not contain density matrices or even
observables and may decay to zero even when mutually decoupled. In
the present paper, we propose a different approach, based primarily
on physical considerations, which yields an inequivalent picture of
open-system adiabaticity that links adiabatic dynamics to the theory
of noiseless subsystems \cite{NS}.

Taking as a ground the basic physical characteristic of adiabatic
closed-system evolutions---namely, that these are quasi-static
evolutions where under sufficiently slow changes of the Hamiltonian
a system in a stationary state evolves so as to remain in a
stationary state with respect to the changed Hamiltonian---we look
for a generalization of this phenomenon to the case of Markovian
dynamics. The key insight of our approach is that the natural
generalization of the eigenspaces of the Hamiltonian corresponding
to distinct eigenvalues are \textit{noiseless subsystems} whose
noiseful cofactors support unique fixed states. A decomposition of
the Hilbert space into such subsystems arises naturally from the
asymptotic behavior of the Lindblad semigroup \cite{Baumgartner08}.
We define adiabaticity as the regime in which the stationary states
over such a noiseless subsystem and its cofactor remain stationary
with respect to the Lindbladian as it changes. We derive an
adiabatic theorem based on this definition.

To illustrate the utility of our formalism, we propose two
applications. One is a framework for decoherence-assisted
computation in noiseless codes which generalizes the approach of
Beige \textit{et al.} \cite{Beige00} to subsystems and general noise
models. The other is a dissipation-driven approach to holonomic
quantum computation based on adiabatic ``dragging'' of subsystems
\cite{HQCsub} along paths that are generally not achievable by
non-dissipative means.

\textit{Generalization of eigenspaces}.---Our starting point is the
observation that the eigenstates of a Hamiltonian $H$ are the
stationary state vectors of its dynamics. In particular, all
stationary density matrices under the evolution $
{d\rho}/{dt}=-i[H,\rho]$ (we set $\hbar=1$) have the
\textit{direct-sum} form $\rho=\bigoplus_ip_i\rho_i$, $\sum_ip_i=1$,
$p_i\geq 0$, where $\rho_i$ are density matrices over the
eigenspaces $\mathcal{H}_i$ of $H$ corresponding to
\textit{distinct} eigenvalues. In more general quantum processes,
the stationary states are organized as operators over
\textit{noiseless subsystems} tensored with a fixed density matrix
over the corresponding noiseful co-subsystem \cite{Blume-Kohout08}.
Consider a time-homogenous finite-dimensional Markovian dynamics
described by the Lindblad equation \cite{Lindbladeqn}
\begin{gather}
\frac{d\rho}{dt}=-i[H,\rho]+\sum_i(L_i\rho
L_i^{\dagger}-\frac{1}{2}L_i^{\dagger}L_i\rho-\frac{1}{2}\rho
L_i^{\dagger}L_i)\equiv\mathcal{L}\rho,\label{Lind}
\end{gather}
where $L_i$ are Lindblad operators. As shown in
Ref.~\cite{Baumgartner08}, Eq.~\eqref{Lind} induces a decomposition
of the Hilbert space
\begin{gather}
\mathcal{H}=\bigoplus_{ij}\mathcal{H}^A_{ij}\otimes\mathcal{H}^B_j\oplus
\mathcal{K},\label{decomp}
\end{gather}
where $\mathcal{H}^A_{ij}$ are \textit{noiseless} subsystems
\cite{NS}, $\mathcal{H}^B_j$ are \textit{noiseful} subsystems that
support unique fixed states, and $\mathcal{K}$ is a decaying
subspace. More particularly, it was shown that for any initial state
$\rho(0)$, the solution of Eq.~\eqref{Lind} satisfies
\begin{gather}
\exists\{p_k,\rho^A_k\}:
\lim_{t\rightarrow\infty}|\rho(t)-\bigoplus_jp_je^{-iH^A_jt}\rho^A_je^{iH^A_jt}\otimes\varrho^B_j
|=0,
\end{gather}
where $\rho^A_j$ are density matrices on the \textit{unitarily
noiseless} subsystems
$\mathcal{H}^A_j=\bigoplus_i\mathcal{H}^A_{ij}$ evolving under the
Hamiltonians $H^A_j$, $\varrho^B_j$ are \textit{fixed full-support
states} on $\mathcal{H}^B_j$, and $\sum_kp_k=1$, $p_k\geq 0$. The
noiseless subsystems $\mathcal{H}^A_{ij}$ are the eigenspaces of
$H^A_j$. The stationary states have the form
$\rho=\bigoplus_{ij}p_{ij}\rho^A_{ij}\otimes\varrho^B_j,\hspace{0.2cm}\sum_{ij}p_{ij}=1,
p_{ij}\geq 0$, where $\rho^A_{ij}$ are density matrices on
$\mathcal{H}^A_{ij}$. This suggests that the subsystems
$\mathcal{H}^A_{ij}$ whose cofactors $\mathcal{H}^B_{j}$ support
unique fixed states $\varrho^B_j$ can be thought of as the
generalization of eigenspaces corresponding to distinct eigenvalues.

How do we find the decomposition \eqref{decomp} for a given
Lindbladian $\mathcal{L}$? An algorithm for finding the noiseless
subsystems of a completely positive trace-preserving (CPTP) map that
runs in time $O[(\textrm{dim}\mathcal{H})^6]$ was described in
Ref.~\cite{Blume-Kohout08} (see also Ref.~\cite{ChoiKribs}). It is
based on finding the left and right operator eigenspaces
corresponding to the eigenvalue $1$ of the CPTP map. Since
Eq.~\eqref{Lind} is equivalent to the continuous application of an
infinitesimal CPTP map, the same algorithm can be used here (the
eigenvalue 1 of the map translates to eigenvalue $0$ of
$\mathcal{L}$).

Before we introduce adiabaticity for Markovian dynamics, let us
briefly review the closed-system case.

\textit{Adiabaticity in closed systems}.---Consider a time-dependent
Hamiltonian $H(t/T)$ changing along a differentiable curve $H(s)$,
$s\in[0,1]$. Let $\epsilon_i(s)$ be an eigenvalue of $H(s)$ with
multiplicity $m$, and $P_i(s)$ be the (twice-differentiable)
projector on the corresponding eigenspace
$\mathcal{H}_i(s)=P_i(s)\mathcal{H}$. [Note that
$m=\textrm{const}(s)$ implies that $\epsilon_i(s)$ is separated from
the rest of the spectrum by a gap. The adiabatic theorem
has been extended to cases without a gap \cite{AvronElgart}, but in
this paper we restrict to the standard formulation.] The eigenspace
$\mathcal{H}_i(t/T)$ is said to evolve adiabatically under $H(t/T)$
if any state initially in $\mathcal{H}_i(0)$ remains in
$\mathcal{H}_i(t/T)$, $t\in[0,T]$. Equivalently, if we change the
basis via a unitary $U(s)$ so that $P_i$ becomes fixed, in the new
basis the dynamics is driven by the effective Hamiltonian $H'(t/T)=
\tilde{H}(t/T)+\frac{1}{T}V(t/T)$, where
$\tilde{H}(s)=U(s)H(s)U(s)^{\dagger}=\epsilon_i(s)P_i+\tilde{H}^{\perp}_i(s)$
with $\tilde{H}^{\perp}_i(s)$ having support on the orthogonal
complement of $\mathcal{H}_i$, and
$V(s)=i\frac{dU(s)}{ds}U^{\dagger}(s)$. Adiabaticity then refers to
the regime in which any state initially in $\mathcal{H}_i$ remains
in $\mathcal{H}_i$ despite the action of $\frac{1}{T}V(t/T)$. The
adiabatic theorem states \cite{Kato50} that in the limit of large $T$, one approaches perfect adiabaticity
where the states in $\mathcal{H}_i$ evolve via the unitary
$U_{i}(s)=\mathcal{T}\exp\left(-i\int_0^s P_iV(q)P_idq\right)$ where
$\mathcal{T}$ denotes time ordering. The error scales with $T$ as $O(\frac{1}{T\Delta})$, where $\Delta>0$ is a fixed energy scale (e.g., the energy gap).

Note that unlike the ``folk'' adiabatic condition which is known to
be insufficient \cite{controversy}, this theorem (similarly to the
one derived below) is concerned with the scaling of the error as a
function of $T$ for a \textit{fixed} curve $H(s)$.

\textit{Adiabaticity in Markovian dynamics}.---Consider a
time-dependent Lindbladian $\mathcal{L}(t/T)$ changing along a
differentiable curve $\mathcal{L}(s)$, $s\in[0,1]$. For every $s$,
$\mathcal{L}(s)$ induces a decomposition of the Hilbert space
$\mathcal{H}=\bigoplus_{ij}\mathcal{H}^A_{ij}(s)\otimes\mathcal{H}^B_j(s)\oplus
\mathcal{K}(s)$ as explained earlier. Let $\mathcal{H}^A_{kl}(s)$
and $\mathcal{H}^B_{l}(s)$ [$\textrm{dim}\mathcal{H}^A_{kl}(s)=m$,
$\textrm{dim}\mathcal{H}^B_{l}(s)=n$] be subsystems of the type
above, and let $\mathcal{P}_{kl}(s)$
[$\mathcal{P}_{kl}(s)\rho=\textrm{Tr}_{B}\{P^{AB}_{kl}(s)\rho
P^{AB}_{kl}(s)\}\otimes\varrho^B_l(s)$ where $P^{AB}_{kl}(s)$ is the
projector on $\mathcal{H}^A_{kl}(s)\otimes\mathcal{H}^B_l(s)$ and
$\textrm{Tr}_B$ denotes partial trace over $\mathcal{H}^B_l$] be the
(twice-differentiable) superoperator projector on the fixed points
over $\mathcal{H}^A_{kl}(s)\otimes\mathcal{H}^B_{l}(s)$.

\textit{Note.} Similarly to the closed-system case, the assumption
that $\textrm{dim}\mathcal{H}^A_{kl}(s)$ and
$\textrm{dim}\mathcal{H}^B_{l}(s)$ are constant implies an analogue
of the gap condition (see Appendix A).

\textit{Definition.} The noiseless subsystem
$\mathcal{H}^A_{kl}(t/T)$ and its noisy cofactor
$\mathcal{H}^B_{l}(t/T)$ evolve adiabatically under
$\mathcal{L}(t/T)$, if any state over $\mathcal{H}^A_{kl}(0)\otimes
\mathcal{H}^B_l(0)$ of the form
$\rho(0)=\rho(0)^A_{kl}\otimes\varrho^B_l(0)$ evolves to a state
$\rho(t)=\rho(t)^A_{kl}\otimes\varrho^B_l(t/T)$ over
$\mathcal{H}^A_{kl}(t/T)\otimes \mathcal{H}^B_l(t/T)$, $t\in[0,T]$.

As in the case of closed systems, it is convenient to consider a
basis rotated by a unitary $U(s)$, in which $\mathcal{H}^A_{kl}$ and
$\mathcal{H}^B_{l}$ are fixed. In this basis, the master equation is
\begin{gather}
\frac{d{\rho}}{dt}=-\frac{i}{T}[V(t/T),{\rho}]+\widetilde{\mathcal{L}}(t/T){\rho},\label{Lind2}
\end{gather}
where $\widetilde{\mathcal{L}}(s)$ is the Lindbladian with $H(s)$
replaced by $U(s){H}(s)U(s)^{\dagger}$ and $L_i(s)$ by
$U(s){L}_i(s)U(s)^{\dagger}$, and
$V(s)=i\frac{dU(s)}{ds}U(s)^{\dagger}$. (We will not use a different
notation for $\rho$ in this basis but will keep in mind the basis we
are working in.) Adiabaticity then means that any state
$\rho(0)=\rho^A_{kl}(0)\otimes\varrho^B_l(0)$ remains of the form
$\rho(t)=\rho^A_{kl}(t)\otimes\varrho^B_l(t/T)$ despite the
perturbation $\frac{1}{T}V(t/T)$.

\textit{Theorem.} Consider Markovian dynamics satisfying the above
assumptions. In the limit of large $T$, perfect adiabaticity is
approached with an error that scales as
$O(\sqrt{\frac{1}{T\Delta}})$, where $\Delta>0$ is some fixed energy
scale. In the adiabatic limit, the states inside
$\mathcal{H}^A_{kl}$ evolve under the unitary
$U^A_{kl}(s)=\mathcal{T}\exp\left(-i\int_0^s\textrm{Tr}_B\{P^{AB}_{kl}V(q)P^{AB}_{kl}I^A_{kl}\otimes\varrho^B_l(q)\}dq
\right)$.

\textit{Proof.} Let us divide the total time $T$ into $N$ steps,
each of length $\delta t$, $T=N\delta t$. We will take $\delta
t={N}/\Delta$ (hence, $T ={N^2}/\Delta$) such that when
$N\rightarrow\infty$, $\delta t$ is short on the time scale of
change of the Lindbladian but long on the time scale for reaching
the asymptotic regime of the instantaneous Lindbladian. The
differentiability assumptions about $\mathcal{L}(s)$ and
$\mathcal{P}_{kl}(s)$ imply that we can write
$\widetilde{\mathcal{L}}(\frac{t+t'}{T})=\widetilde{\mathcal{L}}(\frac{t}{T})+O(\frac{1}{N})$,
$V(\frac{t+t'}{T})=V(\frac{t}{T})+O(\frac{1}{N})$, $t'\in[0,\delta
t]$. The evolution of the density matrix during a single time step
can then be written
\begin{gather}
\rho(t)\rightarrow \rho(t+\delta t)=\mathcal{T}e^{\int_0^{\delta
t}dt'\widetilde{\mathcal{L}}(\frac{t+t'}{T})}\rho(t) +\label{Expansionsimpl22}\\
\int_0^{\delta t}dt'e^{\widetilde{\mathcal{L}}(\frac{t}{T})(\delta
t-t')}\left(\frac{-i}{T}[V(\frac{t}{T}),e^{\widetilde{\mathcal{L}}(\frac{t}{T})t'}\rho(t)]\right)
+O(\frac{1}{N^2}).\notag
\end{gather}

Assume that the state at time $t$ has the form
\begin{gather}
\rho(t)=\rho_{kl}^A(t)\otimes[\varrho^B_l(\frac{t}{T})+O(\frac{1}{N})]+O(\frac{1}{N^2}).\label{form}
\end{gather}
Then the first term on the right-hand side of
Eq.~\eqref{Expansionsimpl22} is $\mathcal{T}e^{\int_0^{\delta
t}dt'\widetilde{\mathcal{L}}(\frac{t+t'}{T})}\rho(t)=\rho_{kl}^A(t)\otimes(\varrho^B_l(\frac{t}{T})+O(\frac{1}{N}))+O(\frac{1}{N^2})$,
since $\mathcal{H}_{kl}^A$ is noiseless and
$\mathcal{T}e^{\int_0^{\delta
t}dt'\widetilde{\mathcal{L}}(\frac{t+t'}{T})}=e^{\delta
t\widetilde{\mathcal{L}}(\frac{t}{T})}+O(\frac{1}{N})$, so for large
$\delta t$ the state on $\mathcal{H}_{l}^B$ decays to
$\varrho^B_l(\frac{t}{T})+O(\frac{1}{N})$ (see Appendix A for an
exact relation to the decay rate). For the second term, ignoring
errors of order $O(\frac{1}{N^2})$, we can use
$\rho(t)=\rho_{kl}^A(t)\otimes\varrho^B_l(\frac{t}{T})$. But
$e^{\widetilde{\mathcal{L}}(\frac{t}{T})t'}$ leaves $\rho(t)$
invariant, so this term becomes $\frac{-i}{T}\int_0^{\delta
t}dt'e^{\widetilde{\mathcal{L}}(\frac{t}{T})(\delta
t-t')}[V(\frac{t}{T}),\rho_{kl}^A(t)\otimes\varrho^B_l(\frac{t}{T})]$.
Using noiseless-subsystem properties of the Lindbladian
\cite{DFScond,Ore}, in Appendix A we show that this term is equal to
$\frac{-i}{T}\int_0^{\delta
t}dt'e^{\widetilde{\mathcal{L}}(\frac{t}{T})(\delta
t-t')}\mathcal{P}_{kl}[V(\frac{t}{T}),\rho_{kl}^A(t)\otimes\varrho^B_l(\frac{t}{T})]+O(\frac{1}{N^2})$.
But $\widetilde{\mathcal{L}}(s)\mathcal{P}_{kl}=0$, so the integral
yields $\frac{-i\delta
t}{T}\mathcal{P}_{kl}[V(\frac{t}{T}),\rho_{kl}^A(t)\otimes\varrho^B_l(\frac{t}{T})]=
-i\frac{\delta t}{T}[
\textrm{Tr}_B\{P^{AB}_{kl}V(\frac{t}{T})P^{AB}_{kl}I^A_{kl}\otimes\varrho^B_l(\frac{t}{T})\},\rho_{kl}^A(t)
]\otimes\varrho_l^B(\frac{t}{T})$ (the last inequality can be
verified by a simple algebra).

We therefore see that if the initial state is of the form \eqref{form}, it
will remain of this form for all times, up to an error $O(\frac{1}{N})=O(\sqrt{\frac{1}{\Delta T}})$ resulting from the accumulation of the
errors $O(\frac{1}{N^2})$ at every step. Moreover,
we see that the reduced density matrix on $\mathcal{H}^A_{kl}$ satisfies the
difference equation ${\rho^A_{kl}(t+\delta t)-\rho^A_{kl}(t)}=-\frac{i\delta t}{T}[
\textrm{Tr}_B\{P^{AB}_{kl}V(\frac{t}{T})P^{AB}_{kl}I^A_{kl}\otimes\varrho^B_l(\frac{t}{T})\},\rho_{kl}^A(t)]+O(\frac{1}{N^2})$,
which in the limit $N\rightarrow \infty$ yields the differential equation $\frac{\partial}{\partial s}{\rho^A_{kl}(Ts)}=-i[
\textrm{Tr}_B\{P^{AB}_{kl}V(s)P^{AB}_{kl}I^A_{kl}\otimes\varrho^B_l(s)\},\rho_{kl}^A(Ts)]$
describing the effective evolution stated in the theorem.

\textit{Note.} Our theorem includes an adiabatic theorem for closed
systems as a special case. However, the convergence rate stated in
our theorem is weaker than the standard one [the error is
$O(\sqrt{\frac{1}{\Delta T}})$ as opposed $O({\frac{1}{\Delta T}})]$
since our proof captures dissipative cases as well. (In Appendix A,
we describe a natural energy scale $\Delta$ associated with the
curve $\mathcal{L}(s)$, which can be regarded as a generalization of
the minimum energy gap.)

\textit{Decoherence-assisted computation in noiseless
codes}.---Computation in noiseless subsystems requires operations
that keep the information inside the code \cite{KBLW01}. However,
the Hamiltonians that preserve the code in general may be rather
complicated and may not be naturally available in a particular
experimental setup. Thus strategies for achieving \textit{encoded
universality} \cite{Bacon01} by other means are of particular
interest \cite{Cory06}.  An immediate implication of the above
theorem is that for the common case of time-homogenous Markovian
noise with Lindbladian $\mathcal{L}$ [to play the role of
$\widetilde{\mathcal{L}}(t/T)$ in Eq.~\eqref{Lind2}], any
Hamiltonian perturbation $\frac{1}{T}V(t/T)$ acting during
$t\in[0,T]$ would give rise to (possibly non-trivial)
\emph{unitary} evolutions inside the noiseless subsystems
$\mathcal{H}^A_{ij}$ of $\mathcal{L}$ within an arbitrary precision
for sufficiently large $T$. Thus given a set of available
interactions $\{{V}_{\mu}\}$ that can be turned on with variable
strength, for a given subsystem $\mathcal{H}^A_{kl}$ one can produce
the set of effective interactions
\begin{gather}
V^{\textrm{eff}}_{\mu}=\textrm{Tr}_B(P^{AB}_{kl}V_{\mu}P^{AB}_{kl}I^A_{kl}\otimes\varrho^B_l).\label{Veff}
\end{gather}
(Note that preparation of the states on $\mathcal{H}^B_l$ is not
needed as they quickly decay to the fixed points.) Encoded
universality is achieved if the set $\{V^{\textrm{eff}}_{\mu}\}$
spans the Lie algebra ${su}(m)$ over $\mathcal{H}^A_{kl}$.
Remarkably this is possible even if the Hamiltonians $\{{V}_{\mu}\}$
commute (see example below).

Such an approach was first proposed in Ref.~\cite{Beige00} for
noiseless subspaces ($\textrm{dim}\mathcal{H}^B_l=1$) under certain
noise models that can be interpreted as continuous Zeno measurements
projecting onto the subspace. Equation \eqref{Veff} provides a
generalization of this idea to noiseless subsystems (that may exist
even when no noiseless subspaces exist) and arbitrary
time-homogenous Markovian models. As an example, in Appendix B we
study a \textit{two-level} noiseless subsystem of three
spin-$\frac{1}{2}$ particles under collective decoherence \cite{NS}.
The noiseless subsystem involves highly entangled states, and
non-local interactions are in principle required to perform
operations on the encoded qubit. However, we find that the
decoherence process itself can be used to induce an effective
universal set of gates on the code by acting with local
Hamiltonians.

\textit{Holonomic quantum computation via dissipation.}---In the
previous method, we assumed that the perturbation $\frac{1}{T}V(s)$
is applied by the experimenter. However, the conclusions are valid
also if we assume that the description is with respect to an
instantaneous basis of a time-dependent noiseless subsystem
$\mathcal{H}^A_{kl}(s)$ of $\mathcal{L}(s)$, where the perturbation
now arises from the time dependence of the basis. As
$\mathcal{L}(s)$ acts trivially on $\mathcal{H}^A_{kl}(s)$, the
effective transformation in $\mathcal{H}^A_{kl}(s)$ is not of
dynamical origin. Indeed, in the adiabatic limit, an initial state
$\rho^{AB}(0)$ over
$\mathcal{H}^A_{kl}(0)\otimes\mathcal{H}^B_{l}(0)$ transforms via
the superoperator $\lim_{\delta s\rightarrow
0}\mathcal{P}_{kl}(1)\mathcal{P}_{kl}(1-\delta
s)...\mathcal{P}_{kl}(\delta s)\mathcal{P}_{kl}(0)$ which is an
intrinsically geometric quantity defined via the projectors
$\mathcal{P}_{kl}(s)$. But the effective unitary on
$\mathcal{H}^A_{kl}(s)$ depends on the choice of basis for
$\mathcal{H}^A_{kl}(s)$ and is not gauge invariant. However, if
$\mathcal{H}^A_{kl}(s)$ is taken around a loop,
$\mathcal{H}^A_{kl}(0)=\mathcal{H}^A_{kl}(1)$, so that the final
basis is the same as the initial one, the resultant transformation
is a gauge-invariant quantity that generalizes the standard holonomy
associated with parallel transport of Hamiltonian eigenspaces
\cite{WZ84}. We note that the idea of adiabatically ``dragging'' a
subsystem (rather than a subspace) along suitable paths in order to
perform geometric gates inside it has been proposed for the case of
Hamiltonian dynamics as a powerful tool for robust computation
\cite{HQCsub}. However, a subsystem cannot be dragged along an
arbitrary path $\mathcal{H}^A(s)$ by a Hamiltonian since some paths
necessarily give rise to correlations between $\mathcal{H}^A(s)$ and
$\mathcal{H}^B(s)$. This problem does not exist here since the
Lindbladian acting on $\mathcal{H}^B(s)$ severs any such
correlations. (For dissipation-driven holonomies in subspaces, see
Ref.~\cite{Carollo05}.)

The mathematical foundations of these geometric transformations will
be studied elsewhere. Here we show that the method can be used for
universal quantum computation. Consider a two-qubit system
$\mathcal{H}=\mathcal{H}^A\otimes\mathcal{H}^B$ and a depolarizing
Markovian channel acting locally on $\mathcal{H}^B$,
$\frac{d\rho^B}{dt}=\mathcal{L}^B\rho^B=\gamma(\frac{I^B}{2}-\rho^B)$.
Consider the unitary $U(s)=e^{-is(a{\sigma_z^{A}\otimes
{\sigma^{B}_z}}+b \sigma^{A}_x\otimes I^B)}$, where
$\sqrt{a^2+b^2}=2\pi$. The Hamiltonian in the exponent can be easily
diagonalized and one sees that $U(1)=U(0)=I$. Hence, if we change
the Lindbladian via the unitary $U(s)$, we will take the noiseless
subsystem $\mathcal{H}^A$ around a loop
$\mathcal{H}^A(s)\otimes\mathcal{H}^B(s)=U(s)\mathcal{H}^A\otimes\mathcal{H}^B$
with a single-valued basis. According to Eq.~\eqref{Veff} (here
$\varrho^B=\frac{I^B}{2}$), the subsystem will experience the
effective Hamiltonian $b\sigma_{x}^A$ which gives rise to the
transformation $e^{-ib\sigma_x^{A}}$ at the closing of the loop.
Similarly, by exchanging $\sigma_z$ and $\sigma_x$, we can generate
the unitary $e^{-ib\sigma_z^{A}}$. To perform an entangling gate
between two qubits, $A$ and $A'$, we can start with the same
Lindbladian acting on $B$ and rotate it via the unitary
$U(s)=e^{-is(a{\sigma_x^{A'}\otimes\sigma_z^{A}\otimes
{\sigma_z^{B}}}+b \sigma_x^{A'}\otimes\sigma_x^{A}\otimes I^B)}$,
which gives rise to $e^{-ib\sigma_x^{A'}\otimes \sigma_x^{A}}$. This
set of gates is universal.

\textit{Conclusion.}---We introduced a theory of adiabatic Markovian
dynamics that relates the notion of adiabaticity to the theory of
noiseless subsystems. We proved an adiabatic theorem for such
dynamics and proposed two novel methods of quantum information
processing based on it---decoherence-assisted computation in
noiseless subsystems and dissipation-driven holonomic computation---that add to the developing picture of dissipation as a powerful quantum computation primitive \cite{DissComp}. A natural problem for future research would be to find exact bounds on the adiabatic error in Markovian dynamics similar to those obtained for closed systems, e.g., in Ref.~\cite{bounds}.

\textit{Acknowledgments.}---We thank Lorenza Viola for helpful
comments. This work was supported by the Spanish MICINN via the
Ram\'{o}n y Cajal program (JC), contract FIS2008-01236/FIS, and
project QOIT (CONSOLIDER2006-00019), and by the Generalitat de
Catalunya via CIRIT 2005SGR-00994. OO was also supported by the
Foundational Questions Institute (FQXi).

\section{Appendix A: Detailed proof of the adiabatic theorem}

\subsection{Preliminaries}

Before we go in detail through the main steps of the proof, it is
convenient to introduce an energy scale $\Delta$ associated with the
curve $\mathcal{L}(s)$. This quantity can be regarded as a
generalization of the minimal spectral gap of the Hamiltonian from
the case of closed systems, and is a suitable choice in view of
certain later calculations. Although it is not the purpose here,
this energy scale could be useful for deriving exact bounds on the
error and not just its scaling with $T$.

As shown in Ref.~\cite{DFScond}, the subsystem $\mathcal{H}^A_{kl}$
is noiseless under the evolution driven by
$\widetilde{\mathcal{L}}(s)$, if and only if for every $s$ the
Hamiltonian and the Lindblad operators have the block forms
\begin{gather}
\widetilde{H}(s)=\begin{bmatrix} I^A\otimes H^B_1(s)&H_2(s)\\
H_2^{\dagger}(s)&H_3(s)
\end{bmatrix},\\
\widetilde{L}_j(s)=\begin{bmatrix} I^A\otimes L^B_{1j}(s)&L_{2j}(s)\\
0&L_{3j}(s)
\end{bmatrix},
\end{gather}
where the upper-left block corresponds to
$\mathcal{H}^A_{kl}\otimes\mathcal{H}^B_l$, and
\begin{gather}
H_2(s)=-\frac{i}{2}\sum_jI^A\otimes L^{B\dagger}_{1j}(s)L_{2j}(s).
\end{gather}
Then it is not difficult to verify (see also Ref.~\cite{Ore}) that
$\widetilde{\mathcal{L}}(s)$ preserves the subspace $\mathcal{B}_2$
of operators with vanishing lower right block,
\begin{gather}
\mathcal{B}_2=\{\tau\in \mathcal{B}(\mathcal{H})\mid \tau=\begin{bmatrix} \tau_1^{AB}&\tau_2\\
\tau_3&0
\end{bmatrix} \},\label{B2}
\end{gather}
where $\mathcal{B}(\mathcal{H})$ denotes the space of linear
operators over $\mathcal{H}$. It further preserves the subspace of
operators with vanishing lower right and offdiagonal blocks,
\begin{gather}
\mathcal{B}_1=\{\tau\in \mathcal{B}(\mathcal{H})\mid \tau=\begin{bmatrix} \tau_1^{AB}&0\\
0&0
\end{bmatrix} \},\label{B1}
\end{gather}
where it acts as
\begin{gather}
\widetilde{\mathcal{L}}(s)\begin{bmatrix} \tau_1^{AB}&0\\
0&0
\end{bmatrix}=\begin{bmatrix} \mathcal{I}^A\otimes\widetilde{\mathcal{L}}^B(s)\tau_1^{AB}&0\\
0&0
\end{bmatrix},
\end{gather}
where $\widetilde{\mathcal{L}}^B(s)$ is a local Lindbladian with
Hamiltonian $H_1^B(s)$ and Lindblad operators $L_{1j}^B(s)$. Note
that $\widetilde{\mathcal{L}}^B(s)$ has a non-degenerate eigenvalue
$0$ with a corresponding right eigenoperator $\varrho^B_l(s)$, and
all its other eigenvalues have negative real parts, since
$\varrho^B_l(s)$ is an attractive fixed point. By continuity, the
magnitudes of the real parts of the non-zero eigenvalues of
$\widetilde{\mathcal{L}}^B(s)$ have a minimum value in the interval
$s\in[0,1]$. Denote that value by $\Delta_1>0$.

We will also need another quantity, $\Delta_2$, which is the minimum
of the magnitudes of the non-zero eigenvalues of
$\widetilde{\mathcal{L}}(s)\mathcal{P}_2$ in the interval
$s\in[0,1]$, where $\mathcal{P}_2$ is the projector on
$\mathcal{B}_2$ (this minimum also exist by the assumption of
continuity in a closed interval). Then we can define
\begin{gather}
\Delta=\min(\Delta_1,\Delta 2),
\end{gather}
which we will serve as a natural energy scale in our analysis. If
$\mathcal{H}^A_{kl}$ is a noiseless subspace
($\dim\mathcal{H}^B_l=1$), $\Delta_1$ does not exist and
$\Delta=\Delta 2$.

We note that in the case of closed systems, where
$\mathcal{H}^A_{k1}$ is an eigenspace of the Hamiltonian, $\Delta$
is exactly the minimum gap that separates this eigenspace from the
rest of the spectrum. This is because the non-zero eigenvalues of
$\widetilde{\mathcal{L}}(s)\mathcal{P}_2$ are $\pm i
(E_k(s)-E_m(s))$, $m\neq k$, where $E_n$ are the energies of the
eigenspaces $\mathcal{H}^A_{n1}$.

\textit{Remark.} The existence of $\Delta>0$ follows naturally from
continuity and the assumption that $\dim\mathcal{H}^A_{kl}$ and
$\dim\mathcal{H}^B_l$ are fixed during the \textit{closed} interval
$s\in[0,1]$. The condition
$\dim\mathcal{H}^A_{kl}=\textrm{const}(s)$ can be thought of as an
analogue of the closed-system requirement that the eigenspace does
not cross other energy levels. However, it may be possible to relax
the condition $\dim\mathcal{H}^B_l=\textrm{const}(s)$ as long as we
require $\Delta>0$ for the \textit{open} subintervals of $s\in[0,1]$
during which $\dim\mathcal{H}^B_l=\textrm{const}(s)$.

Before we proceed, we will need another observation. By the
definition of $\mathcal{H}^A_{kl}\otimes\mathcal{H}^B_l$, the only
right eigenoperators of $\widetilde{\mathcal{L}}(s)$ with eigenvalue
$0$ inside the subspace $\mathcal{B}_2$ are those with $\tau_2=0$
and $\tau_1^{AB}=\tau_{kl}^{A}\otimes\varrho_l^B(s)$. Denote the
subspace of these operators by $\mathcal{B}_0(s)$
[$\mathcal{B}_0(s)\subset\mathcal{B}_1\subset\mathcal{B}_2$]:
\begin{gather}
\mathcal{B}_0(s)=\{\tau\in \mathcal{B}(\mathcal{H})\mid \tau=\begin{bmatrix} \tau_{kl}^{A}\otimes\varrho_l^B(s)&0\\
0&0
\end{bmatrix}, \tau^A_{kl}\in\mathcal{B}(\mathcal{H}^A_{kl}) \}.
\end{gather}
Let $\mathcal{P}'(s)$ be the projector on the subspace
$\mathcal{B}_2\ominus \mathcal{B}_0(s)$. Then the superoperator
$\widetilde{\mathcal{L}}'(s)\equiv\widetilde{\mathcal{L}}(s)\mathcal{P}'(s)$
over $\mathcal{B}_2\ominus \mathcal{B}_0(s)$ has only non-zero
eigenvalues and is therefore invertible. Its inverse,
$\widetilde{\mathcal{L}}'^{-1}(\frac{t}{T})$, has magnitude which is
bounded by the inverse of $\Delta$,
\begin{gather}
\parallel
\widetilde{\mathcal{L}}'^{-1}\parallel\leq\frac{1}{\Delta}.\label{boundedness}
\end{gather}
The boundedness of this operator will be used at a certain stage of
the proof, and the energy scale $\Delta$ was chosen as described
above since it provides the bound \eqref{boundedness}.

We are now ready to go through the steps of the proof.

\subsection{Main proof}

Let us divide the total time $T$ into $N$ time steps, each of length
$\delta t$, $T=N\delta t$. We will take $\delta t={N}/\Delta$
(hence, $T ={N^2}/\Delta$) such that when $N\rightarrow\infty$,
$\delta t$ is short on the time scale of change of the Lidbladian
but long on the time scale for reaching the asymptotic regime of the
instantaneous Lindbladian. The evolution of the density matrix of
the system during one time step can be written
\begin{gather}
\rho(t)\rightarrow \rho(t+\delta t)=\mathcal{T}e_{}^{\int_0^{\delta
t}dt'\widetilde{\mathcal{L}}(\frac{t+t'}{T})}\rho(t) +\notag\\
\int_0^{\delta t}dt''\mathcal{T}e^{\int_{t''}^{\delta
t}dt'''\widetilde{\mathcal{L}}(\frac{t+t'''}{T})}\left(\frac{-i}{T}[V(\frac{t+t''}{T}),\cdot]\right)\notag\\
\times
\mathcal{T}e^{\int_0^{t''}dt'\widetilde{\mathcal{L}}(\frac{t+t'}{T})}\rho(t)
+O(\frac{1}{N^2}),\label{expan}
\end{gather}
where $[V,\cdot]\rho=[V,\rho]$. Our assumption that $\mathcal{L}(s)$
is differentiable and $\mathcal{P}_{kl}(s)$ is twice-differentiable
implies that $\widetilde{\mathcal{L}}(s)$ and $V(s)$ are
differentiable. Hence we have
$\widetilde{\mathcal{L}}(\frac{t+t'}{T})=\widetilde{\mathcal{L}}(\frac{t}{T})+O(\frac{1}{N})$,
$V(\frac{t+t'}{T})=V(\frac{t}{T})+O(\frac{1}{N})$, $t'\in[0,\delta
t]$. We can therefore simplify Eq.~\eqref{expan}:

\begin{gather}
\rho(t)\rightarrow \rho(t+\delta t)=\mathcal{T}e^{\int_0^{\delta
t}dt'\widetilde{\mathcal{L}}(\frac{t+t'}{T})}\rho(t) +\label{Expansionsimpl}\\
\int_0^{\delta t}dt'e^{\widetilde{\mathcal{L}}(\frac{t}{T})(\delta
t-t')}\left(\frac{-i}{T}[V(\frac{t}{T}),\cdot]\right)e^{\widetilde{\mathcal{L}}(\frac{t}{T})t'}\rho(t)
+O(\frac{1}{N^2}).\notag
\end{gather}

Let us now assume that the state at time $t$ has the form
\begin{gather}
\rho(t)=\rho_{kl}^A(t)\otimes[\varrho^B_l(\frac{t}{T})+O(\frac{1}{N})]+O(\frac{1}{N^2}).
\end{gather}
Then the first term on the right-hand side of
Eq.~\eqref{Expansionsimpl} is
\begin{gather}
\mathcal{T}e^{\int_0^{\delta
t}dt'\widetilde{\mathcal{L}}(\frac{t+t'}{T})}\rho(t)=\rho_{kl}^A(t)\otimes\tau^B_l(t)+O(\frac{1}{N^2})
\end{gather}
for some $\tau^B_l(t)$, because $\mathcal{H}_{kl}^A$ is noiseless
under $\widetilde{\mathcal{L}}(s)$. We will now show that for
sufficiently large $\delta t$,
\begin{gather}
\tau^B_l(t)=\varrho^B_l(\frac{t}{T})+O(\frac{1}{N}).
\end{gather}
First of all, we have $\mathcal{T}e^{\int_0^{\delta
t}dt'\widetilde{\mathcal{L}}(\frac{t+t'}{T})}=e^{\delta
t\widetilde{\mathcal{L}}(\frac{t}{T})}+O(\frac{1}{N})$. Also, under
the action of $\widetilde{\mathcal{L}}(\frac{t}{T})$ (for fixed $t$)
any state $\rho^{AB}$ on $\mathcal{H}^A_{kl}\otimes\mathcal{H}^B_l$
decays towards $\rho^A\otimes\varrho^B_l(\frac{t}{T})$, where
$\rho^A=\tr_B\rho^{AB}$, with a rate at least $\Delta$. Since
$\delta t={N}/{\Delta}$, we have $e^{\delta
t\widetilde{\mathcal{L}}(\frac{t}{T})}\rho(t)=\rho_{kl}^A(t)\otimes[\varrho^B_l(\frac{t}{T})+O(\frac{1}{N})]$.
(In fact, the error $O(\frac{1}{N})$ in the latter expression is an
overestimate, but for our purposes this precision suffices.)
Therefore, for the first term on the right-hand side of
Eq.~\eqref{Expansionsimpl} we obtain $\mathcal{T}e^{\int_0^{\delta
t}dt'\widetilde{\mathcal{L}}(\frac{t+t'}{T})}\rho(t)=\rho_{kl}^A(t)\otimes[\varrho^B_l(\frac{t}{T})+O(\frac{1}{N})]+O(\frac{1}{N^2})$.

Next, consider the second term on the right-hand side of
Eq.~\eqref{Expansionsimpl}. Ignoring terms of order
$O(\frac{1}{N^2})$, for this term we can take
$\rho(t)=\rho_{kl}^A(t)\otimes\varrho^B_l(\frac{t}{T})$. The
superoperator $e^{\widetilde{\mathcal{L}}(\frac{t}{T})t'}$ leaves
$\rho(t)$ invariant, so the expression becomes
\begin{gather}
\frac{-i}{T}\int_0^{\delta
t}dt'e^{\widetilde{\mathcal{L}}(\frac{t}{T})(\delta
t-t')}[V(\frac{t}{T}),\rho_{kl}^A(t)\otimes\varrho^B_l(\frac{t}{T})].\label{secondterm}
\end{gather}
We are now going to show that this expression equals
\begin{gather}
\frac{-i}{T}\int_0^{\delta
t}dt'e^{\widetilde{\mathcal{L}}(\frac{t}{T})(\delta
t-t')}\mathcal{P}_{kl}[V(\frac{t}{T}),\rho_{kl}^A(t)\otimes\varrho^B_l(\frac{t}{T})]+O(\frac{1}{N^2}).
\end{gather}
Indeed, let us add and subtract
$\mathcal{P}_{kl}[V(\frac{t}{T}),\rho_{kl}^A(t)\otimes\varrho^B_l(\frac{t}{T})]$
from the operator
$[V(\frac{t}{T}),\rho_{kl}^A(t)\otimes\varrho^B_l(\frac{t}{T})]$ in
expression \eqref{secondterm}. We obtain
\begin{gather}
\frac{-i}{T}\int_0^{\delta
t}dt'e^{\widetilde{\mathcal{L}}(\frac{t}{T})(\delta
t-t')}\mathcal{P}_{kl}[V(\frac{t}{T}),\rho_{kl}^A(t)\otimes\varrho^B_l(\frac{t}{T})]\notag\\
-\frac{i}{T}\int_0^{\delta
t}dt'e^{\widetilde{\mathcal{L}}(\frac{t}{T})(\delta
t-t')}W(t),\label{secondterm2}
\end{gather}
where
\begin{gather}
W(t)=
[V(\frac{t}{T}),\rho_{kl}^A(t)\otimes\varrho^B_l(\frac{t}{T})]\notag\\-\mathcal{P}_{kl}[V(\frac{t}{T}),\rho_{kl}^A(t)\otimes\varrho^B_l(\frac{t}{T})].\notag
\end{gather}
Note that
$[V(\frac{t}{T}),\rho_{kl}^A(t)\otimes\varrho^B_l(\frac{t}{T})]$
belongs to $\mathcal{B}_2$, and therefore $W(t)\in
\mathcal{B}_2\ominus\mathcal{B}_0(\frac{t}{T})$ since $W(t)$ decays
to zero under the action of $\widetilde{\mathcal{L}}(\frac{t}{T})$,
i.e., it has no component in $\mathcal{B}_0(\frac{t}{T})$. We can
therefore formally solve
\begin{gather}
\frac{i}{T}\int_0^{\delta
t}dt'e^{\widetilde{\mathcal{L}}(\frac{t}{T})(\delta
t-t')}W(t)\notag\\
=-\frac{i}{T}\widetilde{\mathcal{L}}'^{-1}(\frac{t}{T})\left(1-e^{\widetilde{\mathcal{L}}(\frac{t}{T})\delta
t}\right)W(t),\label{25}
\end{gather}
where $\widetilde{\mathcal{L}}'^{-1}(\frac{t}{T})$ is the
pseudo-inverse of $\widetilde{\mathcal{L}}(\frac{t}{T})$ over
$\mathcal{B}_2\ominus\mathcal{B}_0(\frac{t}{T})$. According to
Eq.~\eqref{boundedness}, $\parallel
\widetilde{\mathcal{L}}'^{-1}\parallel\leq\frac{1}{\Delta}$, hence
the magnitude of the term in Eq.~\eqref{25} is
$O(\frac{1}{T\Delta})=O(\frac{1}{N^2})$. The only non-trivial
contribution to the expression \eqref{secondterm2} then comes from
\begin{gather}
\frac{-i}{T}\int_0^{\delta
t}dt'e^{\widetilde{\mathcal{L}}(\frac{t}{T})(\delta
t-t')}\mathcal{P}_{kl}[V(\frac{t}{T}),\rho_{kl}^A(t)\otimes\varrho^B_l(\frac{t}{T})]\\=
\frac{-i\delta
t}{T}\mathcal{P}_{kl}[V(\frac{t}{T}),\rho_{kl}^A(t)\otimes\varrho^B_l(\frac{t}{T})]\notag\\
=-i\frac{\delta t}{T}[
\textrm{Tr}_B\{P^{AB}_{kl}V(\frac{t}{T})P^{AB}_{kl}I^A_{kl}\otimes\varrho^B_l(\frac{t}{T})\},\rho_{kl}^A(t)
]\otimes\varrho_l^B(\frac{t}{T}),\notag
\end{gather}
where in the first equality we used that
$\widetilde{\mathcal{L}}(s)\mathcal{P}_{kl}=0$, and the second
equality can be verified by a simple algebra.

We now see that if we start with
$\rho(0)=\rho^A_{kl}(0)\otimes[\varrho^B_l(0)+O(\frac{1}{N})]$, the
state will remain of this form for all times, up to an error of
order $O(\frac{1}{N})=O(\sqrt{\frac{1}{T\Delta}})$ that results from
the accumulation of the errors $O(\frac{1}{N^2})$ at every step
(there are a total of $N$ steps). Moreover, the reduced density
matrix in $\mathcal{H}^A_{kl}$ satisfies the difference equation
\begin{gather}
{\rho^A_{kl}(t+\delta t)-\rho^A_{kl}(t)}=\\-\frac{i\delta t}{T}[
\textrm{Tr}_B\{P^{AB}_{kl}V(\frac{t}{T})P^{AB}_{kl}I^A_{kl}\otimes\varrho^B_l(\frac{t}{T})\},\rho_{kl}^A(t)]+O(\frac{1}{N^2}),\notag
\end{gather}
which in the limit $N\rightarrow \infty$ yields the differential
equation
\begin{gather}
\frac{\partial}{\partial s}{\rho^A_{kl}(Ts)}=\label{diffeq}-i[
\textrm{Tr}_B\{P^{AB}_{kl}V(s)P^{AB}_{kl}I^A_{kl}\otimes\varrho^B_l(s)\},\rho_{kl}^A(Ts)]
\end{gather}
describing the effective evolution stated in the theorem. This
completes the proof.

\section{Appendix B: Example of decoherence-assisted computation in noiseless subsystems}

To illustrate the idea of decoherence-assisted quantum computation
in noiseless subsystems, we consider as an example a
\textit{two-level} noiseless subsystem of three spin-$\frac{1}{2}$
particles under collective decoherence \cite{NS}.

Under the evolution \cite{BP}
\begin{gather}
\frac{d\rho}{dt}=-i\omega[J_z,\rho]+\gamma^-(J_-\rho
J_+-\frac{1}{2}J_+J_-\rho-\frac{1}{2}\rho J_+J_-)\notag\\
+\gamma^+(J_+\rho J_--\frac{1}{2}J_-J_+\rho-\frac{1}{2}\rho
J_-J_+),\label{collective}
\end{gather}
where $J_z=\sum_i\frac{1}{2}\sigma_z^i$ and
$J_{\pm}=\sum_i\sigma^i_{\pm}$ are collective spin operators and
$\gamma^+,\gamma^->0$, there are no non-trivial noiseless subspaces
but there is a noiseless subsystem. The ($\dagger$-closed) operator
algebra generated by $J_{\alpha}$ is isomorphic to
$\mathcal{M}=\bigoplus_{J=1/2}^{3/2}I_{n_J}\otimes
\mathcal{M}(d_J)$, where $J$ is the total angular momentum and
$\mathcal{M}(d_J)$ are $d_J\times d_J$ complex matrix algebras with
multiplicity $n_J$. In particular, $d_J=2J+1$ and $n_{1/2}=2$,
$n_{3/2}=1$. The Hilbert space correspondingly decomposes as
\begin{gather}
\mathcal{H}=\mathcal{H}^A\otimes\mathcal{H}^B\oplus\mathcal{H}^C,
\end{gather}
where $\mathcal{H}^A$ is a noiseless qubit subsystem arising from
the two-fold multiplicity of $\mathcal{M}(2)$, $\mathcal{H}^B$ is
the noisy cofactor supporting $\mathcal{M}(2)$, and $\mathcal{H}^C$
is a noisy subspace supporting $\mathcal{M}(4)$. The subsystems
$\mathcal{H}^A\otimes\mathcal{H}^B$ can be described in the basis
\begin{eqnarray}
|0\rangle^A|0\rangle^B&=&\frac{1}{\sqrt{2}}(|011\rangle-|101\rangle),\\
|0\rangle^A|1\rangle^B&=&\frac{1}{\sqrt{2}}(|010\rangle-|100\rangle),\\
|1\rangle^A|0\rangle^B&=&\frac{1}{\sqrt{6}}(2|110\rangle-|101\rangle-|011\rangle),\\
|1\rangle^A|1\rangle^B&=&\frac{1}{\sqrt{6}}(-2|001\rangle+|010\rangle+|100\rangle).
\end{eqnarray}
One can verify that under the evolution \eqref{collective}, there is
a unique fixed point on $\mathcal{H}^B$,
\begin{gather}
\varrho^B=\frac{\gamma^+}{\gamma^-+\gamma^+}|0\rangle\langle
0|^B+\frac{\gamma^-}{\gamma^-+\gamma^+}|1\rangle\langle 1|^B.
\end{gather}
Similarly, there is a unique fixed point on the subspace
$\mathcal{H}^C$. Using that
\begin{gather}
V^{\textrm{eff}}_{\mu}=\textrm{Tr}_B(P^{AB}_{ij}V_{\mu}P^{AB}_{ij}I^A_{ij}\otimes\varrho^B_j),\label{Veff}
\end{gather}
we obtain that the local Hamiltonian $\sigma_z^1$ gives rise to the
effective Hamiltonian
$\frac{\gamma^--\gamma^+}{2\sqrt{3}(\gamma^-+\gamma^+)}{\sigma_x^A}$,
where ${\sigma_x^A}=|0\rangle\langle 1|^A+|1\rangle\langle 0|^A$ is
the encoded Pauli operator $\sigma_x$ on $\mathcal{H}^A$. Similarly,
$\sigma_z^3$ gives rise to
$\frac{2(\gamma^--\gamma^+)}{{3}(\gamma^-+\gamma^+)}{\sigma_z^A}$.
These two Hamiltonians generate $SU(2)$ on $\mathcal{H}^A$.

For universal computation one needs the ability to entangle multiple
noiseless qubits, e.g., by bringing different blocks together
\cite{KBLW01} and manipulating the logical information inside the
resulting larger noiseless subsystems. This problem can be treated
via the same approach.

\end{document}